%% file: main.tex
\documentclass[10pt,twocolumn,letterpaper]{article}

\usepackage[pagenumbers]{cvpr}   
\usepackage[table]{xcolor}
\usepackage{booktabs}
\usepackage{graphicx}
\usepackage[ruled,vlined]{algorithm2e}
\usepackage[accsupp]{axessibility}  
\input{preamble}

\definecolor{cvprblue}{rgb}{0.21,0.49,0.74}
\usepackage[pagebackref,breaklinks,colorlinks,allcolors=cvprblue]{hyperref}

\def\paperID{*****} 
\def\confName{CVPR}
\def\confYear{2026}

\title{HIVE: Query, Hypothesize, Verify — An LLM Framework for Multimodal Reasoning-Intensive Retrieval}

\author{
Mahmoud Abdalla$^{1}$ \qquad
Mahmoud SalahEldin Kasem$^{1}$ \qquad
Mohamed Mahmoud$^{1}$ \qquad \\
Mostafa Farouk Senussi$^{1}$ \qquad
Abdelrahman Abdallah$^{2}$ \qquad
Hyun-Soo Kang$^{1}$\thanks{Corresponding author.}
\\[0.5em]
$^{1}$ Chungbuk National University
$^{2}$ University of Innsbruck\\[0.5em]
}

\begin{document}
\maketitle
\input{sec/0_abstract}    
\input{sec/1_intro}
\input{sec/2_relatedwork}
\input{sec/3_model}

\input{sec/4_Experiments}
\input{sec/6_Conclusion}

{
    \small
    \bibliographystyle{ieeenat_fullname}
    \bibliography{main}
}


\end{document}

%% file: preamble.tex
\usepackage{booktabs}
\usepackage{multirow}
\usepackage{array}



%% file: sec/0_abstract.tex
\begin{abstract}
Multimodal retrieval models fail on reasoning-intensive queries where images 
(diagrams, charts, screenshots) must be deeply integrated with text to identify 
relevant documents --- the best multimodal model achieves only 27.6 nDCG@10 
on MM-BRIGHT, underperforming even strong text-only retrievers (32.2). 
We introduce \textbf{HIVE} (\textbf{H}ypothesis-driven \textbf{I}terative 
\textbf{V}isual \textbf{E}vidence Retrieval), a plug-and-play framework that injects explicit visual–text reasoning into a retriever via LLMs. 
HIVE operates in four stages: (1) initial retrieval over the corpus, 
(2) LLM-based compensatory query synthesis that explicitly articulates visual 
and logical gaps observed in top-$k$ candidates, (3) secondary retrieval with 
the refined query, and (4) LLM verification and reranking over the union of 
candidates. Evaluated on the multimodal-to-text track of MM-BRIGHT (2,803 
real-world queries across 29 technical domains), HIVE achieves a new 
state-of-the-art aggregated nDCG@10 of \textbf{41.7} --- a \textbf{+9.5} 
point gain over the best text-only model (DiVeR: 32.2) and \textbf{+14.1} 
over the best multimodal model (Nomic-Vision: 27.6), where our 
reasoning-enhanced base retriever contributes 33.2 and the HIVE framework 
adds a further \textbf{+8.5} points --- with particularly strong results 
in visually demanding domains (Gaming: 68.2, Chemistry: 42.5, 
Sustainability: 49.4).
Compatible with both standard and reasoning-enhanced retrievers, HIVE 
demonstrates that LLM-mediated visual hypothesis generation and verification 
can substantially close the multimodal reasoning gap in retrieval.\footnote{\url{https://github.com/mm-bright/multimodal-reasoning-retrieval}}
\end{abstract}

%% file: sec/1_intro.tex
\section{Introduction}

\begin{figure*}[t]
    \centering
    \includegraphics[width=0.9\linewidth]{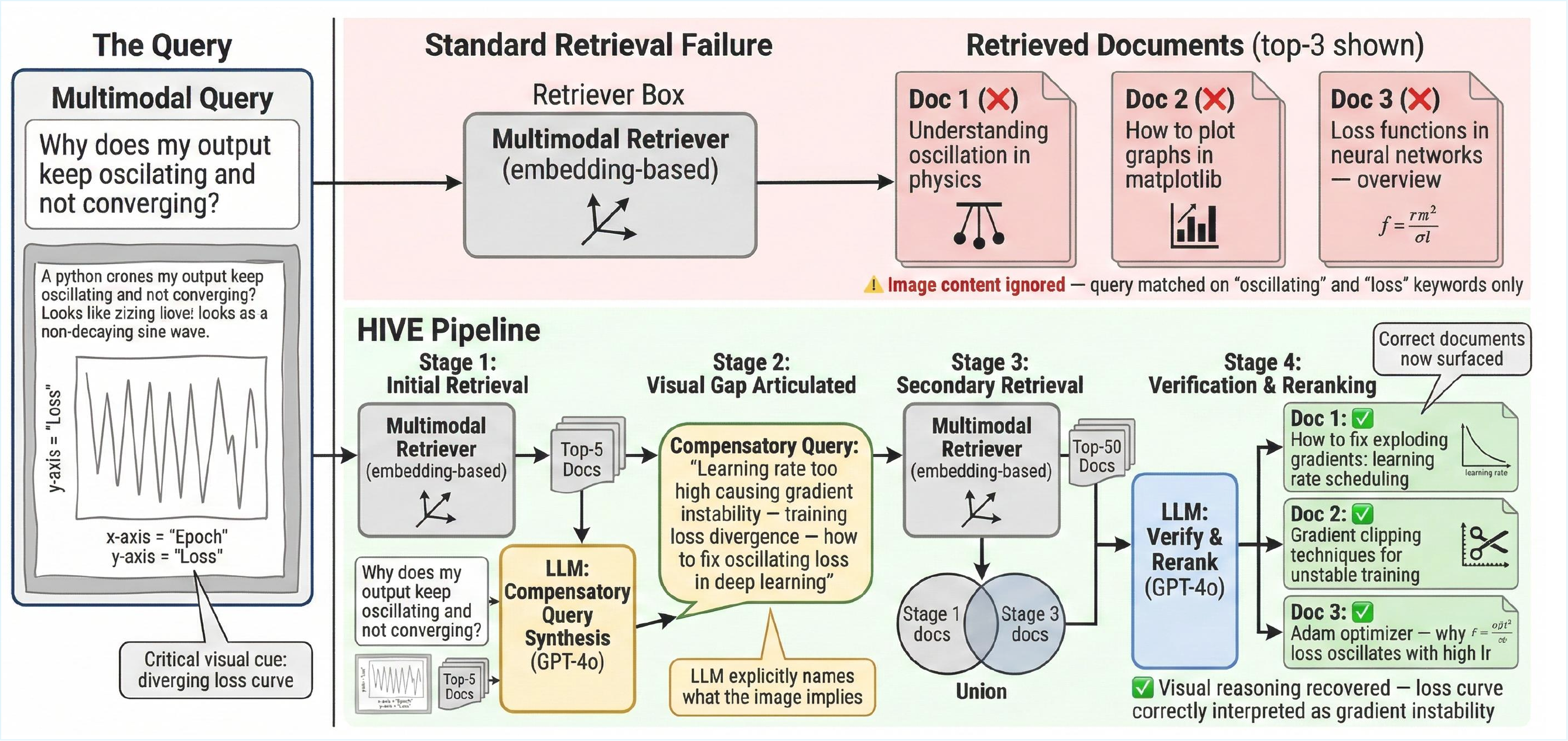}
    \caption{An example where standard multimodal retrievers fail to identify 
    the relevant document because the query image (a circuit diagram) contains 
    critical visual cues that text-only or embedding-based matching cannot capture. 
    HIVE generates a compensatory query that explicitly articulates these visual 
    gaps, enabling successful retrieval.}
    \label{fig:example}
\end{figure*}

The ability to retrieve relevant information from large corpora is fundamental 
to knowledge-intensive applications such as question answering, retrieval-augmented 
generation, and agentic systems~\cite{lewis2020rag}. Recent advances 
in dense retrieval have yielded powerful embedding models that capture rich semantic 
relationships between queries and documents~\cite{karpukhin2020dpr}. 
However, these models are predominantly designed for text-only retrieval, where 
surface-level semantic matching is often sufficient. Real-world queries are increasingly multimodal. Users post screenshots of error 
messages, attach diagrams from scientific papers, or include charts from financial 
reports when seeking help online. In these settings, retrieval requires \textit{reasoning} 
— understanding what the image depicts, how it relates to the text query, and which 
documents in the corpus collectively address both dimensions. As shown in Figure~\ref{fig:example}, 
a query containing a circuit diagram and the text ``why is my LED not lighting up?'' 
cannot be resolved by matching keywords alone; the retrieval system must reason about 
the circuit topology to identify the relevant document. Recent vision and vision-language models demonstrate strong task-specific reasoning, such as rationale-driven anomaly detection~\cite{abdalla2025think}, industrial defect classification~\cite{kasem2025attention}, and video understanding~\cite{mahmoud2025two}. However, these approaches are designed for closed-set prediction and do not address how visual reasoning can be leveraged to improve open-domain retrieval over large text corpora.

This challenge is starkly reflected in recent benchmarks. On MM-BRIGHT~\cite{abdallah2026mmbright}, 
the first multimodal benchmark for reasoning-intensive retrieval, state-of-the-art 
multimodal models achieve only 27.6 nDCG@10 — \textit{lower} than the best 
text-only retriever (32.2). Adding visual information actively hurts performance, 
revealing a fundamental gap: current multimodal retrievers embed images and text 
jointly but lack the capacity to reason about what visual content \textit{implies} 
for document relevance. A natural question arises: \textit{can we inject visual reasoning into retrieval 
without retraining any model?} Existing approaches to improve retrieval quality 
rely on query expansion~\cite{wang2023query2doc}, chain-of-thought 
prompting~\cite{wei2022chain}, or reranking~\cite{sun2023rankgpt,abdallah2025dear} — but these 
operate purely in the text domain and ignore the visual gap. On the other hand, 
fine-tuning multimodal retrievers requires large labeled datasets and is 
computationally prohibitive for most practitioners.

We introduce \textbf{HIVE} (\textbf{H}ypothesis-driven \textbf{I}terative \textbf{V}isual \textbf{E}vidence Retrieval), a plug-and-play framework that 
addresses this gap by harnessing LLMs as explicit visual reasoners within the 
retrieval pipeline. Crucially, HIVE requires no additional training beyond the 
base retrieval model --- the framework operates entirely at inference time, 
making it compatible with any retriever, from off-the-shelf embedding models 
to reasoning-enhanced fine-tuned retrievers. HIVE is motivated by a key insight: LLMs, when provided with top-$k$ retrieved 
documents and a description of the query image, can identify what is missing and 
generate a precise compensatory query that captures the visual reasoning the base 
retriever could not perform. This is analogous to how a human expert, upon seeing 
an unsatisfying set of search results, reformulates their query by explicitly 
articulating what they are looking for — except HIVE does this with full awareness 
of both the image content and the retrieved documents.

Our contributions are as follows:
\begin{itemize}
    \item We identify and formalize the \textit{visual reasoning gap} in 
    multimodal retrieval: the failure of embedding-based models to reason about 
    what query images imply for document relevance (\S\ref{sec:main_results}).
    
    \item We propose \textbf{HIVE}, a plug-and-play, model-agnostic framework 
    that injects LLM-driven visual hypothesis generation and verification into 
    any base retriever via a four-stage pipeline (\S\ref{sec:method}).
    
    \item We demonstrate that HIVE achieves a new state-of-the-art aggregated 
    nDCG@10 of \textbf{41.7} on MM-BRIGHT's multimodal-to-text track, 
    outperforming the best multimodal model by \textbf{+14.1} points and the 
    best text-only model by \textbf{+9.5} points, with consistent gains across 
    all 29 domains (\S\ref{sec:domain}).
\end{itemize}

%% file: sec/2_relatedwork.tex
\section{Related Work}

\subsection{Dense Retrieval and Reasoning-Intensive Benchmarks}

Dense retrieval using bi-encoder models has become the dominant paradigm for 
large-scale information retrieval~\cite{karpukhin2020dpr, reimers2019sentence,ali2025sustainableqa}. These models independently encode queries and documents into a 
shared embedding space and retrieve via efficient nearest-neighbor 
search~\cite{johnson2019faiss}. While highly effective on fact-seeking queries 
in benchmarks such as BEIR~\cite{thakur2021beir} and MTEB~\cite{muennighoff2023mteb}, 
bi-encoders struggle when relevance requires multi-step reasoning rather than 
surface-level semantic matching. The BRIGHT and Tempo benchmark~\cite{su2025bright,abdallah2026tempo} exposed 
this limitation clearly: even the strongest embedding models (nDCG@10 of 59.0 on BEIR) 
collapse to 18.3 on reasoning-intensive queries. Subsequent work has sought to address 
this gap through reasoning-aware fine-tuning~\cite{shao2025reasonir,das2025rader}, 
iterative query expansion~\cite{wang2023query2doc, lei2025thinkqe}, and LLM-based 
reranking~\cite{sun2023rankgpt,abdallah2025rankify,mozafari2025good, zhuang2025rankr1, weller2025rank1}. DIVER~\cite{long2025diver} 
integrates all three components — document preprocessing, feedback-based query 
expansion, and hybrid pointwise-listwise reranking — achieving state-of-the-art 
results on BRIGHT. 

\subsection{Multimodal Embedding Models}

Contrastive vision-language models such as CLIP~\cite{radford2021clip} and 
SigLIP~\cite{zhai2023siglip} established the foundation for joint image-text 
embedding by aligning visual and textual representations through large-scale 
contrastive pretraining. However, these models produce a single shared embedding 
space and lack the capacity for fine-grained, instruction-following retrieval. 
Nomic Embed Vision~\cite{nomic2024vision} addressed this by sharing an embedding 
space between a vision encoder and a strong text model, enabling zero-shot 
image-text retrieval with competitive performance on standard benchmarks.

Recent MLLMs-based retrievers --- VLM2Vec~\cite{jiang2024vlm2vec}, 
GME~\cite{zhang2024gme}, and Qwen3-VL-Embedding~\cite{li2026qwen3vl} --- 
achieve strong results through contrastive training on multimodal benchmarks. 
Despite these advances, all share a fundamental limitation: embedding-based 
similarity cannot reason about what a query image \textit{implies} for 
document relevance. MM-BRIGHT confirmed this directly: the best multimodal 
model (Nomic-Vision: 27.6) underperforms the best text-only retriever 
(DiVeR: 32.2), showing visual reasoning cannot be reduced to embedding fusion.

\subsection{Visual Document Retrieval}

ColPali~\cite{faysse2024colpali} treats document pages as images and embeds 
them via a VLM using ColBERT-style late interaction, bypassing OCR entirely. 
DSE~\cite{ma2024dse} and VisRAG~\cite{yu2024visrag} similarly embed full 
page images for dense retrieval. While effective for visual document 
retrieval, these methods assume visual queries \textit{and} documents; 
HIVE addresses multimodal-to-text retrieval where documents are purely 
textual.


\subsection{LLM-Augmented Query Reformulation}

Query expansion and reformulation have long been used to bridge the vocabulary gap between queries and documents~\cite{robertson2009bm25}. With the advent of LLMs, several works have proposed generating hypothetical documents~\cite{gao2022hyde}, chain-of-thought reasoning queries~\cite{su2025bright}, answer scent queries~\cite{abdallah2025asrank}, or iteratively expanded queries~\cite{lei2025thinkqe, long2025diver} to improve the retrieval of reasoning-intensive content. Re-Invoke~\cite{chen2024reinvoke} applies LLMs to unsupervised tool retrieval by enriching tool documents offline and extracting user intent at inference time. RankGPT~\cite{sun2023rankgpt} demonstrated that LLMs can directly rerank retrieved passages through sliding window prompting, while Rank1~\cite{weller2025rank1} and RankR1~\cite{zhuang2025rankr1} further improved reranking via reasoning optimized LLMs.


\subsection{Multimodal Reasoning-Intensive Retrieval}

MM-BRIGHT~\cite{abdallah2026mmbright} introduced the first benchmark for 
reasoning-intensive multimodal retrieval, spanning 2,803 queries across 29 
technical domains with four task types of increasing complexity. The benchmark 
revealed that existing multimodal models systematically fail on queries requiring 
visual reasoning, with the best multimodal model underperforming text-only 
baselines. Concurrent work MR$^2$-Bench~\cite{mr2bench2025} proposed a 
complementary benchmark emphasizing visual-centric reasoning, including spatial 
puzzles and multi-image relational tasks, and found that reasoning-enhanced 
reranking strategies yield consistent gains. Our work directly addresses the 
challenge identified by these benchmarks: HIVE is the first framework to 
explicitly model the visual reasoning gap at the retrieval level, achieving 
a new state-of-the-art on MM-BRIGHT's multimodal-to-text track.

%% file: sec/3_model.tex
\begin{figure*}[t]
    \centering
    \includegraphics[width=\textwidth]{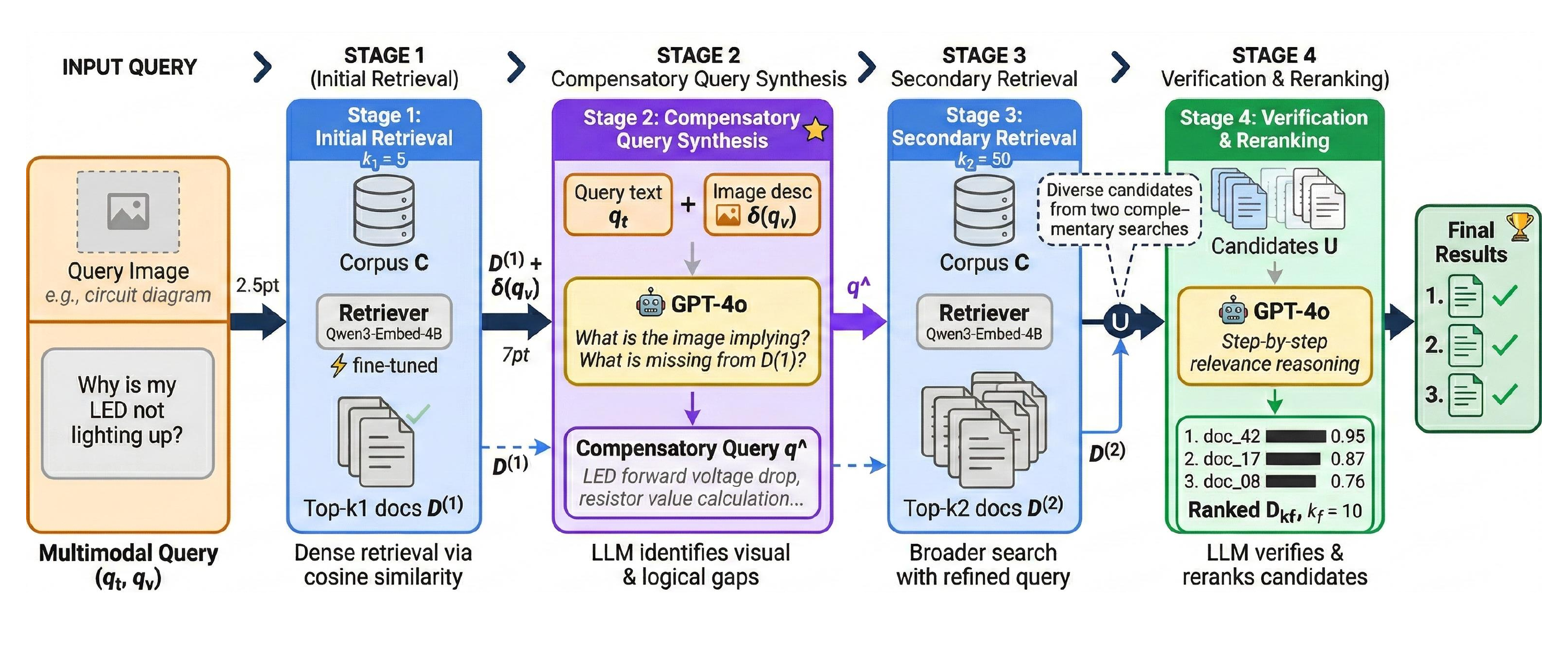}
    \caption{
        Overview of the HIVE framework. Given a multimodal query $(q_t, q_v)$, 
        HIVE operates in four stages: 
        \textbf{(1) Initial Retrieval} --- a base retriever retrieves 
        a small probe set $\mathcal{D}^{(1)}$ of top-$k_1$ candidates; 
        \textbf{(2) Compensatory Query Synthesis} --- an LLM inspects the probe 
        documents alongside the image description $\delta(q_v)$ to identify 
        visual and logical gaps, generating a targeted compensatory query $\hat{q}$; 
        \textbf{(3) Secondary Retrieval} --- the base retriever uses $\hat{q}$ 
        to retrieve a broader candidate set $\mathcal{D}^{(2)}$; and 
        \textbf{(4) Verification \& Reranking} --- an LLM re-evaluates the 
        union $\mathcal{U} = \mathcal{D}^{(1)} \cup \mathcal{D}^{(2)}$ against 
        the original multimodal query and produces the final ranked list 
        $\hat{\mathcal{D}}_{k_f}$. No additional training is required beyond 
        the base retriever.
    }
    \label{fig:hive_overview}
\end{figure*}
\section{Method}
\label{sec:method}
\subsection{Problem Formulation}

We address the \textit{multimodal-to-text} retrieval task, in which each query 
consists of a text component and one or more associated images, while the 
document corpus contains text-only passages. Formally, let $\mathcal{C} = 
\{d_1, d_2, \ldots, d_N\}$ denote a corpus of $N$ text documents. A multimodal 
query is a pair $q = (q_t, q_v)$, where $q_t$ is the textual component and 
$q_v$ is an image (e.g., a diagram, chart, or screenshot). The objective is to 
retrieve an ordered set of documents $\hat{\mathcal{D}}_k \subset \mathcal{C}$, 
$|\hat{\mathcal{D}}_k| = k$, that are most relevant to the full intent expressed 
by $(q_t, q_v)$.

Standard dense retrievers operate by encoding query and documents into a shared 
embedding space and ranking by cosine similarity:
\begin{equation}
    \text{score}(q, d) = \text{cos}(\phi(q_t, q_v),\ \psi(d))
\end{equation}
where $\phi(\cdot)$ and $\psi(\cdot)$ denote query and document encoders 
respectively. While effective for surface-level matching, this formulation 
provides no mechanism for reasoning about what $q_v$ \textit{implies} for 
document relevance --- the core challenge in reasoning-intensive multimodal 
retrieval. HIVE addresses this by introducing an LLM-mediated intermediate stage 
that explicitly constructs a \textit{compensatory query} $\hat{q}$ capturing 
visual and logical cues missing from the initial retrieval, and then verifies 
and reranks the union of candidates from both rounds.

\subsection{Overview of HIVE}

HIVE is a plug-and-play, four-stage retrieval framework that augments any base 
retriever with LLM-driven visual hypothesis generation and verification. The 
term \textit{iterative} reflects that retrieval is performed in two successive 
passes, each conditioned on the output of the previous stage --- an iterative 
refinement over the candidate set rather than a single-pass lookup. An 
overview is shown in Figure~\ref{fig:hive_overview}. Given a multimodal query 
$(q_t, q_v)$ and a text corpus $\mathcal{C}$, HIVE proceeds as follows:

\begin{enumerate}
    \item \textbf{Initial Retrieval.} A base retriever $R$ retrieves the top-$k_1$ 
    candidate documents $\mathcal{D}^{(1)} = R(q_t, q_v, \mathcal{C}, k_1)$.
    
    \item \textbf{Compensatory Query Synthesis.} An LLM inspects $\mathcal{D}^{(1)}$ 
    alongside the image description $\delta(q_v)$ and generates a compensatory 
    query $\hat{q}$ that articulates the visual and logical gaps in the initial results.
    
    \item \textbf{Secondary Retrieval.} The same retriever $R$ uses $\hat{q}$ to 
    retrieve a broader set of candidates $\mathcal{D}^{(2)} = R(\hat{q}, q_v, 
    \mathcal{C}, k_2)$, where $k_2 \gg k_1$.
    
    \item \textbf{Verification and Reranking.} An LLM re-evaluates the union 
    $\mathcal{U} = \mathcal{D}^{(1)} \cup \mathcal{D}^{(2)}$ against the original 
    query $(q_t, q_v)$ and produces a final ranked list $\hat{\mathcal{D}}_{k_f}$.
\end{enumerate}



\begin{table*}[t]
\centering
\resizebox{0.87\textwidth}{!}{%
\begin{tabular}{l c c c c c c c c}
\toprule
\textbf{DOMAIN} & \textbf{BGE-VL} & \textbf{CLIP} & \textbf{GME-2B} & 
\textbf{GME-7B} & \textbf{JINA-CLIP} & \textbf{NOMIC} & \textbf{SIGLIP} & 
\textbf{HIVE} \\
\midrule
\rowcolor{teal!20} \multicolumn{9}{c}{\textit{STEM \& Life Sciences}} \\
\midrule
Acad      & 4.2  & 4.8  & 16.2 & 27.6 & 22.3 & 22.6 & 3.6  & \textbf{45.4} \\
Bio       & 5.7  & 14.8 & 22.9 & 15.2 & 20.5 & 26.9 & 11.9 & \textbf{42.7} \\
Chem      & 10.8 & 9.6  & 27.2 & 21.9 & 30.6 & 30.6 & 11.6 & \textbf{42.5} \\
Phys      & 6.8  & 6.1  & 13.3 & 14.0 & 14.4 & 17.2 & 7.3  & \textbf{25.1} \\
Math      & 13.1 & 17.9 & 16.4 & 9.3  & 27.0 & 34.0 & 15.3 & \textbf{45.2} \\
Earth     & 10.1 & 10.9 & 20.5 & 26.2 & 24.6 & 30.1 & 11.8 & \textbf{44.6} \\
BioAc     & 13.3 & 11.4 & 10.5 & 13.4 & 19.4 & 23.4 & 14.8 & \textbf{40.1} \\
BioInf    & 11.6 & 9.4  & 21.1 & 19.2 & 23.7 & 33.8 & 16.8 & \textbf{34.8} \\
Med       & 12.6 & 9.8  & 22.7 & 19.0 & 26.8 & 33.9 & 9.1  & \textbf{42.0} \\
\midrule
\rowcolor{teal!20} \multicolumn{9}{c}{\textit{Software \& Technical Systems}} \\
\midrule
Apple     & 7.2  & 12.3 & 23.9 & 17.0 & 24.3 & 28.7 & 4.4  & \textbf{36.1} \\
Ubuntu    & 11.6 & 5.5  & 25.9 & 34.2 & 26.1 & 34.3 & 12.6 & \textbf{53.6} \\
BTC       & 8.9  & 8.3  & 18.2 & 19.6 & 22.6 & 22.7 & 10.0 & \textbf{37.3} \\
Crypto    & 11.3 & 14.8 & 9.8  & 7.1  & 15.5 & 22.4 & 10.2 & \textbf{22.5} \\
QC        & 4.5  & 2.6  & 5.9  & 5.6  & 10.8 & 12.1 & 2.6  & \textbf{14.9} \\
Robot     & 16.1 & 10.6 & 15.8 & 18.7 & 19.0 & 30.3 & 14.3 & \textbf{39.1} \\
Sales     & 14.2 & 2.3  & 31.1 & \textbf{47.3} & 32.3 & 26.2 & 6.5  & 44.9 \\
\midrule
\rowcolor{teal!20} \multicolumn{9}{c}{\textit{Social Sciences \& Humanities}} \\
\midrule
Econ      & 9.5  & 6.0  & 10.0 & 12.6 & 13.5 & 21.1 & 9.8  & \textbf{49.2} \\
Psych     & 6.4  & 8.7  & 15.6 & 18.6 & 20.8 & 23.9 & 7.9  & \textbf{41.9} \\
Phil      & 2.4  & 5.4  & 15.2 & 18.0 & 19.4 & 21.7 & 7.0  & \textbf{30.1} \\
Law       & 10.2 & 19.7 & 30.7 & 35.0 & 35.3 & 47.6 & 16.4 & \textbf{64.6} \\
Christ    & 8.9  & 15.0 & 20.0 & 26.5 & 21.0 & 30.9 & 13.0 & \textbf{49.2} \\
Islam     & 12.0 & 10.7 & 25.8 & 32.0 & 24.3 & 28.9 & 6.5  & \textbf{41.0} \\
\midrule
\rowcolor{teal!20} \multicolumn{9}{c}{\textit{Applied Domains}} \\
\midrule
Aviat     & 9.6  & 15.4 & 16.2 & 17.0 & 24.3 & 24.1 & 9.2  & \textbf{40.1} \\
Game      & 17.5 & 19.1 & 41.6 & 43.9 & 45.6 & 43.1 & 21.4 & \textbf{68.2} \\
GIS       & 13.8 & 13.1 & 15.5 & 15.6 & 20.3 & 25.8 & 16.5 & \textbf{36.5} \\
PM        & 8.6  & 8.9  & 21.9 & 33.2 & 20.5 & 27.6 & 12.4 & \textbf{41.9} \\
Sustain   & 10.1 & 9.0  & 16.7 & 25.6 & 24.3 & 24.7 & 11.5 & \textbf{49.4} \\
Travel    & 10.1 & 16.1 & 23.9 & 30.8 & 26.6 & 36.7 & 13.1 & \textbf{51.8} \\
Quant     & 8.1  & 2.1  & 12.4 & 15.3 & 11.6 & 16.2 & 5.8  & \textbf{34.2} \\
\midrule
\textbf{Average} & 10.0 & 10.4 & 19.5 & 22.0 & 23.0 & 27.6 & 10.8 & \textbf{41.7} \\
\bottomrule
\end{tabular}%
}
\caption{Per-domain nDCG@10 on MM-BRIGHT (multimodal-to-text track) across 
all 29 domains, grouped by thematic category. HIVE achieves the highest 
average score and the best score in 28 of 29 domains. Best scores per domain 
in \textbf{bold}. Average is computed as a query-weighted macro average 
following the official MM-BRIGHT evaluation 
protocol~\cite{abdallah2026mmbright}, where domains with more queries 
contribute proportionally; per-domain scores are rounded to one decimal place 
after aggregation.}
\label{tab:mmbright_results}
\end{table*}
\subsection{Stage 1: Base Retriever}
\label{sec:retriever}

HIVE is retriever-agnostic and can wrap any dense retrieval model. In our 
experiments, we use \textbf{HIVE-Retriever}, a reasoning-enhanced embedding 
model independently fine-tuned from Qwen3-Embedding-4B on synthetic 
hard-negative contrastive data spanning medical, mathematical, and general 
domains. Our Retriever encodes both text and image-text queries into a shared 
embedding space using the hidden state of the \texttt{[EOS]} token from the 
last layer, and retrieves via cosine similarity:
\begin{equation}
    \text{score}(q, d_i) = \cos\bigl(E(q_t, q_v),\ E(d_i)\bigr)
\end{equation}

For multimodal queries, we represent $q_v$ through its image caption $\delta(q_v)$ generated by GPT-4o, concatenated with $q_t$ as the query input. This caption is used consistently across all four stages of HIVE. We note that GPT-4o is thus invoked in three roles: (1) image captioning to produce $\delta(q_v)$, (2) compensatory query synthesis in Stage~2, and (3) verification and reranking in Stage~4. All three calls are at inference time and require no gradient updates beyond the base retriever.

\subsection{Stage 2: Compensatory Query Synthesis}
\label{sec:hypothesis}

The central hypothesis of HIVE is that the failure of initial retrieval on 
reasoning-intensive queries is systematic: the top-$k_1$ retrieved documents 
collectively reveal which visual or logical dimensions the query requires but 
the retriever failed to surface. An LLM, given both the image description 
$\delta(q_v)$ and $\mathcal{D}^{(1)}$, is uniquely positioned to identify 
these gaps and synthesize a targeted compensatory query.

Formally, let $\mathcal{D}^{(1)} = \{d_{r_1}, \ldots, d_{r_{k_1}}\}$ be the 
initial probe results ordered by retrieval score. We construct the following 
prompt to the LLM:
\begin{equation}
    \hat{q} = \mathcal{L}\bigl(\textsc{HypothesisPrompt}(q_t, \delta(q_v), 
    \{d_{r_i}\}_{i=1}^{k_1})\bigr)
\end{equation}

The \textsc{HypothesisPrompt} instructs the LLM to: (1) identify what the 
image depicts beyond what the text query states, (2) assess which aspects of 
the query intent are missing or only partially addressed by the probe documents, 
and (3) produce a concise compensatory query $\hat{q}$ that bridges these gaps. 

\paragraph{Design rationale.} Unlike standard query expansion methods that 
reformulate the text query in isolation~\cite{wang2023query2doc, lei2025thinkqe}, 
the compensatory query synthesis in HIVE explicitly conditions on: (i) the 
visual content of the query image, and (ii) the actual content of the top-$k_1$ 
retrieved documents. This dual conditioning allows HIVE to generate targeted 
expansions that address specifically what is missing, rather than general 
paraphrases of the original query. For example, given a query image showing a 
Python stack trace and retrieved documents discussing general debugging practices, 
HIVE would generate a compensatory query articulating the specific error type 
and module visible in the image.

\subsection{Stage 3: Secondary Retrieval}
\label{sec:secondary}

Given the compensatory query $\hat{q}$, we perform a second retrieval pass with 
a larger budget $k_2 \gg k_1$:
\begin{equation}
    \mathcal{D}^{(2)} = R(\hat{q}, q_v, \mathcal{C}, k_2)
\end{equation}

We use $k_2 = 50$ and $k_1 = 5$ by default (ablated in 
Section~\ref{sec:ablation}). The larger $k_2$ is intentional: since $\hat{q}$ 
targets a different semantic subspace than the original query, a broader search 
increases coverage of relevant documents that the initial pass missed. The final 
candidate pool is the union of both passes:
\begin{equation}
    \mathcal{U} = \mathcal{D}^{(1)} \cup \mathcal{D}^{(2)}, \quad 
    |\mathcal{U}| \leq k_1 + k_2
\end{equation}
where duplicates are removed while preserving retrieval order.

\subsection{Stage 4: Verification and Reranking}
\label{sec:reranking}

Given $\mathcal{U}$, a final LLM-based verification and reranking step produces 
the output list $\hat{\mathcal{D}}_{k_f}$. The LLM is provided with the full 
union of candidates alongside the original query $(q_t, \delta(q_v))$ and asked 
to reason step-by-step about relevance before producing a ranked list of the 
top-$k_f$ document IDs:
\begin{equation}
    \hat{\mathcal{D}}_{k_f} = \mathcal{L}\bigl(\textsc{VerifyPrompt}(q_t, 
    \delta(q_v), \mathcal{U}, k_f)\bigr)
\end{equation}

The \textsc{VerifyPrompt} instructs the LLM to act as a verification agent: 
given the full visual-textual query context and all candidate documents, reason 
through which documents most effectively resolve the user's multimodal intent, 
then output a ranked JSON list of document IDs. Documents not appearing in the 
LLM's top-$k_f$ list are assigned residual scores based on their original 
retrieval rank, ensuring full coverage for downstream evaluation. The final 
scores are assigned as:
\begin{equation}
    \text{score}(d_i) = \begin{cases}
        S_{\max} - \text{rank}_i & \text{if } d_i \in \hat{\mathcal{D}}_{k_f} \\
        S_{\text{base}} - \text{offset}_i & \text{otherwise}
    \end{cases}
\end{equation}
where $S_{\max} = 1000$ and $S_{\text{base}} = 500$ are fixed constants and 
$\text{offset}_i$ reflects the position in the unranked residual list.

%% file: sec/4_Experiments.tex
\section{Experiments}

\subsection{Dataset}

We evaluate HIVE on MM-BRIGHT~\cite{abdallah2026mmbright}, the first 
reasoning-intensive multimodal retrieval benchmark. MM-BRIGHT consists of 
\textbf{2,803 queries} spanning \textbf{29 technical domains}, including 
Gaming, Chemistry, Law, Sustainability, Earth Science, Mathematics, Computer 
Science, Medicine, and others. Each query is a multimodal pair $(q_t, q_v)$ 
comprising a text question and one or more associated images (diagrams, charts, 
screenshots, molecular structures, etc.), paired with a text-only document corpus.



\subsection{Experimental Setup}

\paragraph{Computational Infrastructure.}
The base retriever was trained on \textbf{4$\times$ NVIDIA H100 80GB 
GPUs} using distributed data-parallel training. HIVE inference requires no 
additional GPU training; all LLM calls for compensatory query synthesis 
(Stage~2) and verification and reranking (Stage~4) use \textbf{GPT-4o}. Embedding inference for all retrieval stages is performed on a single node.

\paragraph{HIVE Hyperparameters.}
Unless otherwise specified, we use the following default hyperparameters 
throughout all experiments: probe size $k_1 = 5$, secondary retrieval size 
$k_2 = 50$, final reranking output $k_f = 10$, and GPT-4o temperature $= 0$ 
(deterministic decoding). Sensitivity to $k_1$ and $k_2$ is analyzed in 
Section~\ref{sec:ablation}.


\subsection{Baselines}

We compare against two families of baselines, following the MM-BRIGHT 
evaluation protocol.

\paragraph{Multimodal Retrievers.}
These models encode the full multimodal query $(q_t, q_v)$:
\begin{itemize}
    \item \textbf{CLIP}~\cite{radford2021clip}: Contrastive image-text model 
    with shared embedding space.
    \item \textbf{SigLIP}~\cite{zhai2023siglip}: Sigmoid-based contrastive VLM 
    with improved image-text alignment.
    \item \textbf{Jina-CLIP}~\cite{koukounas2024jinaclip}: Multi-task 
    contrastive model supporting text-image and text-text retrieval.
    \item \textbf{Nomic Embed Vision}~\cite{nomic2024vision}: Shares an 
    embedding space between a vision encoder and a strong text model, enabling 
    zero-shot multimodal retrieval.
    \item \textbf{BGE-VL}~\cite{bge2024vl}: Multimodal embedding model from 
    the BGE family supporting fused-modal retrieval.
    \item \textbf{GME-Qwen2-VL}~\cite{zhang2024gme}: Universal multimodal 
    embedding model built on Qwen2-VL, supporting any-to-any retrieval across 
    text, image, and fused modalities.
\end{itemize}

\paragraph{Text-Only Retrievers.}
These models encode only the text component $q_t$ of the query:
\begin{itemize}
    \item \textbf{BM25}: Sparse lexical baseline using BM25 scoring.
    \item \textbf{E5-Mistral-7B}~\cite{wang2023e5}: Instruction-tuned dense 
    retriever built on Mistral-7B.
    \item \textbf{GTE-Qwen2-7B}~\cite{li2023gte}: General text embedding model 
    from the GTE family.
    \item \textbf{DiVeR}~\cite{long2025diver}: Reasoning-enhanced retriever 
    fine-tuned on Qwen3-Embedding-4B with synthetic hard-negative data and 
    iterative query expansion; current state-of-the-art on BRIGHT.
\end{itemize}

\section{Results}

\subsection{Main Results.}
\label{sec:main_results}
\begin{figure}[t]
    \centering
    \includegraphics[width=\linewidth]{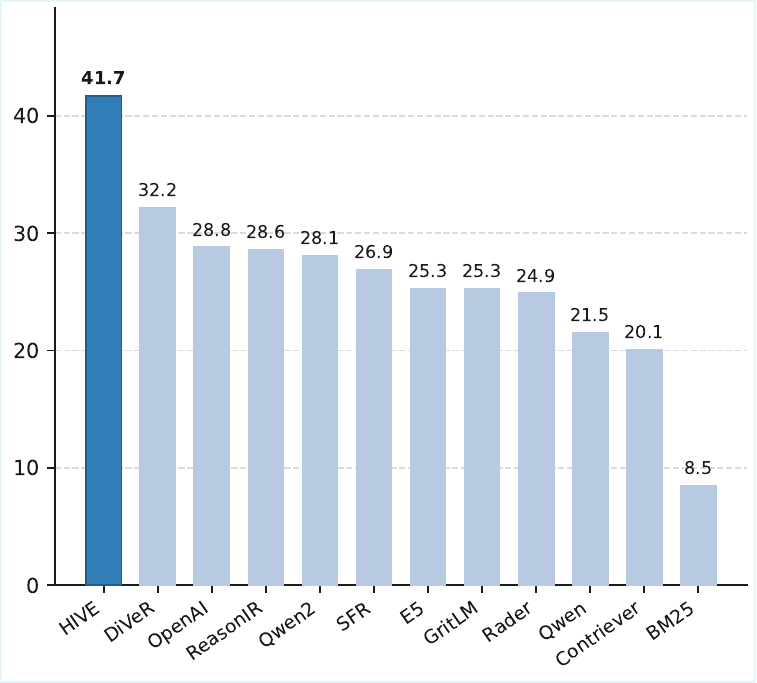}
    \caption{Average nDCG@10 on MM-BRIGHT (multimodal-to-text track) for 
    all evaluated models. HIVE achieves 41.7, outperforming the best 
    multimodal baseline (Nomic-Vision: 27.6) by +14.1 points.}
    \label{fig:hive_comparison}
\end{figure}

Figure~\ref{fig:hive_comparison} and Table~\ref{tab:mmbright_results} present 
the aggregated and per-domain nDCG@10 results across all 29 MM-BRIGHT domains. 
HIVE achieves a new state-of-the-art aggregated nDCG@10 of \textbf{41.7}, 
surpassing the strongest multimodal baseline (Nomic-Vision: 27.6) by 
\textbf{+14.1 points} and the best text-only model (DiVeR: 32.2) by 
\textbf{+9.5 points}. We attribute these gains to two complementary 
components: our reasoning-enhanced base retriever (HIVE-Retriever: 33.2, 
already surpassing DiVeR by +1.0), and the HIVE inference framework which 
contributes an additional \textbf{+8.5 points} (33.2~$\to$~41.7) with no 
additional training beyond the base retriever.

The performance gap between HIVE and all baselines is consistent and 
substantial. The second-best model, Nomic-Vision (27.6), is followed by 
Jina-CLIP (23.0), GME-7B (22.0), GME-2B (19.5), SigLIP (10.8), CLIP (10.4), 
and BGE-VL (10.0). Notably, all contrastive VLM baselines score below 28, 
confirming the finding from MM-BRIGHT~\cite{abdallah2026mmbright} that 
standard embedding-based similarity is fundamentally insufficient for 
reasoning-intensive multimodal retrieval.

\subsection{Domain-Level Analysis}
\label{sec:domain}
Table~\ref{tab:mmbright_results} reports per-domain nDCG@10 across all 29 MM-BRIGHT domains, grouped into four thematic categories. HIVE achieves the best performance in \textbf{28 of 29 domains}, demonstrating that the gains are not driven by a few outlier domains but reflect a consistent and systematic improvement across the full benchmark. The single exception is Salesforce (Sales), where GME-7B (47.3) narrowly outperforms HIVE (44.9) --- a domain characterized by highly structured CRM screenshots whose visual layout GME-7B's contrastive training specifically captures, reducing the marginal benefit of compensatory query synthesis.

Several patterns emerge from the domain-level breakdown. First, HIVE shows 
the strongest absolute scores in \textbf{Gaming} (68.2), \textbf{Law} (64.6), 
\textbf{Ubuntu} (53.6), \textbf{Travel} (51.8), and \textbf{Economics} (49.2). 
Second, the largest relative gains over Nomic-Vision occur in visually demanding 
domains such as \textbf{Academic} (+22.8), \textbf{Economics} (+28.1), and 
\textbf{Sustainability} (+24.7), where diagrams and charts carry dense 
domain-specific information that embedding similarity cannot capture. Third, even in domains where Nomic-Vision is relatively strong (Law: 47.6, 
BioInformatics: 33.8), HIVE still achieves improvements (+17.0 and +1.0 
respectively). The marginal gain in BioInformatics (+1.0) is the single 
exception and is explained by domain characteristics: bioinformatics queries 
predominantly feature sequence-based images (e.g., alignment plots, 
phylogenetic trees) whose visual content carries limited additional semantic 
signal beyond what the text query already expresses. In such cases, GPT-4o's 
image description $\delta(q_v)$ adds little new information to the 
compensatory query, reducing HIVE's advantage to near-zero. This pattern 
is consistent with Quantum Computing (14.9) and Cryptography (22.5), where 
visual content is similarly abstract --- confirming that HIVE's gains are 
largest when images carry dense, domain-specific information not expressible 
in the text query alone. The weakest HIVE performance is observed in 
\textbf{Quantum Computing} (14.9) and \textbf{Cryptography} (22.5) --- both 
highly abstract technical domains where visual content offers limited 
additional signal beyond the text query.

\subsection{Plug-and-Play Analysis.}
To demonstrate that HIVE is a general framework rather than a 
retriever-specific optimization, Table~\ref{tab:plugin} reports nDCG@10 
for each base retriever with and without HIVE. We note that HIVE-Retriever 
(33.2) already outperforms DiVeR (32.2) by +1.0 as a standalone model; 
the HIVE framework then contributes an additional +8.5 points (33.2~→~41.7), 
confirming that the framework's gains are independent of the base retriever's 
strength. HIVE yields consistent improvements across all base retrievers, 
regardless of whether they are contrastive VLMs or reasoning-enhanced text 
models, with larger gains for weaker bases (e.g., +10.1 for CLIP), confirming 
that compensatory queries cover a wider semantic gap when the initial probe 
set is less informative.

\begin{table}[t]
\centering
\caption{HIVE as a plug-and-play framework across base retrievers. 
$\Delta$ denotes absolute nDCG@10 improvement from applying HIVE.}
\label{tab:plugin}
\begin{tabular}{lccc}
\toprule
\textbf{Base Retriever} & \textbf{Base} & \textbf{+HIVE} & $\boldsymbol{\Delta}$ \\
\midrule
CLIP                     & 10.4 & 20.5          & +10.1 \\
GME-Qwen2-VL-2B          & 19.5 & 25.3          & +5.8  \\
GME-Qwen2-VL-7B          & 22.0 & 27.5          & +5.5  \\
Nomic-Vision             & 27.6 & 34.1          & +6.5  \\
HIVE-Retriever           & 33.2 & \textbf{41.7} & +9.5  \\
\bottomrule
\end{tabular}
\end{table}

\subsection{Ablation Study.}
\label{sec:ablation}
Table~\ref{tab:ablation_component} evaluates the contribution of each HIVE 
stage by progressively adding components on top of the HIVE-Retriever base. 
The results confirm that both the compensatory query synthesis (Stages 2--3) 
and the verification and reranking (Stage 4) contribute substantially and 
independently to the final performance, with neither stage alone sufficient 
to achieve the full gain. Applying Stage~4 only (verify/rerank over the 
initial probe set $\mathcal{D}^{(1)}$) yields 37.2, while applying 
Stages~2+3 only (expand without verification) yields 38.1 --- below the 
41.7 achieved by the full pipeline. This confirms a clear synergy: Stage~2--3 
expands coverage by surfacing relevant documents the base retriever missed, 
while Stage~4 filters the noise introduced by the broader secondary retrieval.

\begin{table}[t]
\centering
\caption{Component ablation on MM-BRIGHT (HIVE-Retriever base).}
\label{tab:ablation_component}
\resizebox{0.45\textwidth}{!}{%
\begin{tabular}{lc}
\toprule
\textbf{Configuration} & \textbf{nDCG@10} \\
\midrule
HIVE-Retriever (base, no HIVE Framework)             & 33.2 \\
+ Stage 2 only (compensatory query, no verify)       & 35.3 \\
+ Stage 4 only (verify/rerank $\mathcal{D}^{(1)}$)  & 37.2 \\
+ Stages 2+3 (expand only, no verify)               & 38.1 \\
\textbf{+ Stages 1--4 (full HIVE)}                  & \textbf{41.7} \\
\bottomrule
\end{tabular}}
\end{table}

\begin{figure}[t]
    \centering
    \includegraphics[width=\linewidth]{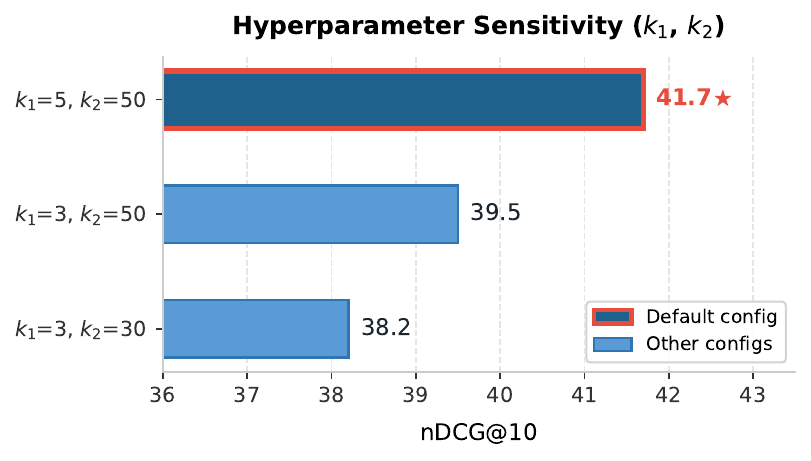}
    \caption{Hyperparameter sensitivity on MM-BRIGHT. Each bar shows 
    nDCG@10 for a $(k_1, k_2)$ configuration. The default setting 
    ($k_1{=}5$, $k_2{=}50$) achieves the best performance of 41.7, 
    while even the lightest configuration ($k_1{=}3$, $k_2{=}30$) 
    yields a substantial +5.0 gain over the base retriever (33.2).}
    \label{fig:hyperparam}
\end{figure}

Figure~\ref{fig:hyperparam} further reports sensitivity to the probe 
size $k_1$ and secondary retrieval size $k_2$. Performance improves 
consistently as both $k_1$ and $k_2$ increase, with $k_1{=}5$, $k_2{=}50$ 
offering the best tradeoff between retrieval coverage and LLM context length. 
Smaller probe sets ($k_1{=}3$, $k_2{=}30$) already yield a 5-point 
improvement over the base retriever (38.2 vs.\ 33.2), confirming that 
even a light application of HIVE is beneficial.


%% file: sec/6_Conclusion.tex
\section{Conclusion}

We presented HIVE (Hypothesis-driven Iterative Visual Evidence Retrieval), 
a plug-and-play framework that addresses the fundamental limitation of 
existing multimodal retrievers on reasoning-intensive queries. While state-of-the-art multimodal embedding models score below 28 nDCG@10 on MM-BRIGHT --- and are outperformed by text-only retrievers --- HIVE achieves 41.7 nDCG@10 through two complementary contributions: a reasoning-enhanced base retriever (33.2) and the HIVE inference framework (+8.5 points), which injects explicit visual reasoning via LLM-mediated compensatory query synthesis and verification at inference time. The key insight driving HIVE is that retrieval failure on multimodal 
reasoning queries is systematic and diagnosable: an LLM, given the query 
image description and the top-$k$ probe documents, can reliably identify 
what visual and logical cues are missing and synthesize a targeted 
compensatory query to recover them. This two-pass architecture --- initial 
retrieval, hypothesis generation, secondary retrieval, and verification --- 
requires no additional training beyond the base retriever and is compatible 
with any dense retrieval backbone, as confirmed by consistent gains across 
five base retrievers spanning contrastive VLMs and reasoning-enhanced text 
models.

Domain-level analysis across all 29 MM-BRIGHT domains reveals that HIVE's gains are largest in visually demanding domains where images carry dense structured information (gains over Nomic-Vision --- Gaming: +25.1, Economics: +28.1, Sustainability: +24.7), while gains are more modest in abstract domains 
with limited visual signal (Quantum Computing: 14.9, Cryptography: 22.5). 
This pattern provides a clear signal for future work: the bottleneck in 
abstract domains lies in GPT-4o's ability to ground ambiguous or 
low-resolution images, suggesting that visual grounding confidence 
estimation could further improve HIVE's effectiveness. Future directions include extending HIVE to multi-image and video-to-text retrieval, replacing GPT-4o with smaller open-source models to reduce cost, and learning compensatory query generation end-to-end via reinforcement learning from retrieval feedback.

\section*{Acknowledgment}
This work was supported by Innovative Human Resource Development for Local Intellectualization program through the Institute of Information \& Communications Technology Planning \& Evaluation(IITP) grant funded by the Korea government(MSIT) (IITP-2026-RS-2020-II201462, 50\%), and partly supported by the National Research Foundation of Korea (NRF) grant funded by the Korea government (Ministry of Science and ICT) (RS-2023-NR076833)